\documentclass{vgtc}                          % final 
\ifpdf%                                % if we use pdflatex
  \pdfoutput=1\relax                   % create PDFs from pdfLaTeX
  \usepackage{graphicx}                % allow us to embed graphics files
  \DeclareGraphicsExtensions{.pdf,.png,.jpg,.jpeg} % for pdflatex we expect .pdf, .png, or .jpg files
\else%                                 % else we use pure latex
  \usepackage{graphicx}                % allow us to embed graphics files
  \DeclareGraphicsExtensions{.eps}     % for pure latex we expect eps files
\fi%

\graphicspath{{figures/}{pictures/}{images/}{./}} % where to search for the images
\usepackage{subfigure}
\usepackage{microtype}                 % use micro-typography (slightly more compact, better to read)
\PassOptionsToPackage{warn}{textcomp}  % to address font issues with \textrightarrow
\usepackage{textcomp}                  % use better special symbols
\usepackage{mathptmx}                  % use matching math font
\usepackage{times}                     % we use Times as the main font
\usepackage{cite}                      % needed to automatically sort the references
\usepackage{tabu}                      % only used for the table example
\usepackage{booktabs}                  % only used for the table example
\usepackage{amssymb}
\usepackage{caption}
\usepackage{amsmath,mathtools}
\usepackage{graphicx}
\usepackage{changepage}
\usepackage{ragged2e}
\usepackage{hyperref} 
\usepackage{xcolor}
\newcommand{\norm}[1]{\left\lVert#1\right\rVert}
\onlineid{0}

\vgtccategory{Research}

\vgtcinsertpkg

\title{Rotation-constrained optical see-through headset calibration with bare-hand alignment}

\author{Xue Hu\thanks{e-mail: xue.hu17@imperial.ac.uk}\\ %
        \parbox{1.8in}{\scriptsize \centering Mechatronics in Medicine Lab \\ Imperial College London}
\and Ferdinando Rodriguez y Baena\thanks{e-mail: f.rodriguez@imperial.ac.uk}\\ %
     \parbox{1.8in}{\scriptsize \centering Mechatronics in Medicine Lab \\ Imperial College London}
\and Fabrizio Cutolo\thanks{e-mail: fabrizio.cutolo@endocas.unipi.it}\\ %
     \parbox{2.3in}{\scriptsize \centering Department of Information Engineering \\ University of Pisa}}
\abstract{
The inaccessibility of user-perceived reality remains an open issue in pursuing the accurate calibration of optical see-through (OST) head-mounted displays (HMDs). Manual user alignment is usually required to collect a set of virtual-to-real correspondences, so that a default or an offline display calibration can be updated to account for the user's eye position(s). Current alignment-based calibration procedures usually require point-wise alignments between rendered image point(s) and associated physical landmark(s) of a target calibration tool. As each alignment can only provide one or a few correspondences, repeated alignments are required to ensure calibration quality. 

This work presents an accurate and tool-less online OST calibration method to update an offline-calibrated eye-display model. The user's bare hand is markerlessly tracked by a commercial RGBD camera anchored to the OST headset to generate a user-specific cursor for correspondence collection. The required alignment is object-wise, and can provide thousands of unordered corresponding points in tracked space. The collected correspondences are registered by a proposed rotation-constrained iterative closest point (rcICP) method to optimise the viewpoint-related calibration parameters. We implemented such a method for the Microsoft HoloLens 1. The resiliency of the proposed procedure to noisy data was evaluated through simulated tests and real experiments performed with an eye-replacement camera.
According to the simulation test, the rcICP registration is robust against possible user-induced rotational misalignment. With a single alignment, our method achieves 8.81 arcmin (1.37 mm) positional error and 1.76 $^{\circ}$ rotational error by camera-based tests in the arm-reach distance, and 10.79 arcmin (7.71 pixels) reprojection error by user tests.
} % end of abstract

\CCScatlist{ 
\CCScat{H.5.1}{Information interfaces and presentation}%
 {Multimedia Information Systems};
 {Artificial, Augmented, and Virtual
realities};
  \CCScat{H.5.2}{Information interfaces and presentation}%
{User Interfaces} 
{Ergonomics, Evaluation/methodology, Screen design.}
}

\begin{document}

\firstsection{Introduction}
\maketitle
In visual augmented reality (AR) experience, defining the appropriate spatial location of the computer-generated 3D content with respect to the real scene under observation is the principal factor that provides the user with a sense of perceptual congruity (i.e., locational realism) \cite{GrubertItoh:2018Survey}. Optical see-through (OST) head-mounted displays (HMDs) are at the leading edge of the AR research. In OST devices, the computer-generated virtual image is projected onto a semi-transparent optical combiner (OC) placed in front of the user's eyes, so that the user's pupil can intercept both the light rays coming from the physical environment and those emitted from the microdisplay \cite{sielhorst2008advanced, FerrariParallaxOST:2020}. 
Collimation optics are placed between the microdisplay and the OC so that the virtual 2D image is focused on one or more virtual plane(s) at a comfortable viewing distance \cite{Holliman:3DDisplaysReview}. The almost unaltered direct view of the real world allows for a safe and immersive AR experience \cite{qian2017comparison}. 
However, the inaccessibility of user-perceived retinal images makes OST display calibration particularly challenging \cite{Gilson:2008}. The complexity and unreliability of the calibration procedure required to ensure accurate virtual-to-real alignment is the major obstacle to the widespread adoption of OST HMDs across medical and industrial settings. 

OST calibration aims to estimate the rendering camera's projection parameters that ensure an appropriate alignment between the real target scene perceived in the user's line-of-sight and its virtual homologous rendered on the HMD virtual screen \cite{GrubertItoh:2018Survey}. The eye-display system is usually modelled as an off-axis pinhole camera, the image plane of which corresponds to the see-through virtual screen and the projection centre of which corresponds to the nodal point of the user's eye \cite{ItohIndica:2014}. 
The model contains both hardware-related and human perspective-related contributions. The human perspective can be directly measured by automatic eye-tracking \cite{PlopskiEyeTracking:2015, ItohIndica:2014} or indirectly estimated from manual user alignment \cite{TuceryanSPAAM:2000, NavabEasySPAAM:2004}. Of the two options, alignment-based methods are more viable across commercial HMDs due to their weak requirement on dedicated hardware to track the user's eye(s) \cite{MoserLeapMotion:2016}. Moreover, unlike the automatic methods that track the eyeball centre rather than the actual optical eye centre \cite{GrubertItoh:2018Survey,PlopskiEyeTracking:2015}, alignment-based methods can yield authentic viewpoint-related parameters and are thus more accurate when the eye rotates to focus at different distances \cite{MoserImpactofAlignmentonSPAAM:2108}. 

In alignment-based calibration procedures, users need to visually align on-screen virtual points with real-world targets by observing the world through the OC. The set of associated 2D-3D point correspondences are collected to optimise the unknown parameters required for display update \cite{ABDELAZIZDLT:2015,hartley_zisserman:2004}.
Nevertheless, alignment-based calibration can be highly time-consuming (i.e., multiple alignments are required to yield accurate results \cite{Axholt2011PinholeCC}), tedious (i.e., calibration should be repeated any time the HMD moves on the user's head), and sensitive to the alignment quality performed by the user.
Following the most popular Single-Point Active Alignment Method (SPAAM) \cite{tuceryan2002single}, many alignment-based OST calibrations have been developed \cite{makibuchi2016virc, azimi2017alignment, SunOSTCAlibrationOnlineSurgery:2020}. However, most of these rely on sparse point-wise correspondences collection, specially-made calibration tools, or at least multiple repeated alignments.

In this work, we present an accurate and tool-less online OST calibration method developed upon a homography-corrected off-axis eye-display model \cite{hu2020alignment, cutolo2020off} to account for the viewpoint-related contribution. A commercial RGBD camera anchored to the headset is exploited to markerlessly track the user's bare-hand in real-time. The user's hand is first sampled at an initial position to generate a user-specific contour cursor at the peripersonal location. The cursor is then displayed by the HMD, over which the user needs to align his/her hand. The two dense point clouds, sampled by the RGBD camera at the cursor-generation moment and the alignment moment, can be registered by a proposed rotation constrained-iterative closest point (rcICP) method to optimise the unknown parameters required for the OST display update. The proposed calibration procedure has been implemented with a consumer-level headset, HoloLens 1 (Microsoft Inc.). Our analysis includes both simulated tests and physical experiments based on an eye-replacement camera and actual users. After comparing our method with the state-of-the-art, a final discussion is provided.

The main features of our proposed calibration procedure are highlighted below:
\begin{enumerate}
\item The proposed calibration directly exploits the raw dense point cloud sampled by a commercial RGBD camera for OST-HMD calibration. Our proposed rcICP registration can uniquely utilise the implicit correspondence in such collected data.
\item The proposed calibration is user-centric: the method exploits no tool but the user's bare-hand, the involved alignment is object-wise but not point-wise, the alignment cursor is generated in a user-specific way, and a single alignment can potentially ensure the reliable result.
\item The proposed calibration is robust: the rcICP registration ensures the robustness of our calibration against the rotational real-to-virtual misalignment, which may exist in the data collected by novice users by means of dedicated calibration tools.
% even with poor alignment quality in orientation
\end{enumerate}

\section{Notation}
The following notation is used throughout the paper: coordinate systems are denoted by uppercase letters, such as the 3D world coordinate frame ${W}$. Scalars are denoted by lowercase letters, such as the display focal length $f$. Matrices are denoted by uppercase bold letters. If the matrix presents a rigid transformation, the subscript and superscript indicate the source and destination reference frames. For example, the transformation from coordinate A to B is: $\prescript{B}{A}{\mathbf{T}} = \begin{bmatrix}
\prescript{B}{A}{\mathbf{R}} & \prescript{B}{A}{\mathbf{t}}\\
0&1
\end{bmatrix}$. Vectors are denoted by lowercase bold letters, such as the translation from point A to B expressed in screen coordinates:
$\mathbf{t}_{\scriptscriptstyle{AB}}^{\scriptscriptstyle{(S)}} = [x_{\scriptscriptstyle{AB}},y_{\scriptscriptstyle{AB}},z_{\scriptscriptstyle{AB}}]^{\scriptscriptstyle T}$. 

\section{Related Works}

\subsection{Two-stage OST calibration models}
OST calibration aims to find the projection matrix of the rendering camera $\mathbf{P}$ that can display a 2D pixel $\mathbf{m}$ on the semi-transparent virtual screen so that it aligns with its 3D counterpart $\mathbf{v}^{\scriptscriptstyle {(W)}}$ in the user's retina:
\begin{equation} 
\zeta\mathbf{m} =  \mathbf{P} ~ \mathbf{V} ~ \mathbf{v}^{\scriptscriptstyle {(W)}}
\end{equation}
where $\mathbf{V}$ is the transformation from world to the rendering camera frustum (i.e., the view matrix). $\zeta$ is a scaling factor due to the equivalence between points expressed in homogeneous coordinates. $\zeta$ equals the distance from $\mathbf{v}^{\scriptscriptstyle {(W)}}$ to the rendering camera's principal plane \cite{fusiello2006elements}. $\mathbf{P}$ contains 11 intrinsic and extrinsic degrees of freedom (DOFs) \cite{GrubertItoh:2018Survey}. A highly tedious alignment procedure is required to derive all unknown DOF. For example, SPAAM calibration requires at least six but practically more than 20 alignments for each eye \cite{TuceryanSPAAM:2000}. 

To reduce the required number of online alignments and thus the user workload, a two-phase calibration is commonly adopted in the literature \cite{GrubertItoh:2018Survey}. In the offline phase, the hardware-related projection parameters are determined for an initial generic viewpoint by performing a sort of factory calibration, ideally in a controlled setup. In the online phase, the offline calibration is refined by adjusting a small subset of viewpoint-related parameters to account for the specific user's eye(s) position. Many models have been proposed for the online update of OST calibration. 

\subsection{Environment-centric and user-centric alignment}
% Over the past decade, much research effort has been devoted to improving the manual alignment by reducing the required number of alignment and the amount of physical exertion per alignment.
% Since the introduction of the most successful SPAAM \cite{TuceryanSPAAM:2000}, the user's head was allowed to freely move during alignments by tracking the HMD pose on the fly. The unconstrained head mobility reduces the difficulty and discomfort during alignment, but naturally incurs higher noise in the collected correspondences due to the involuntary head motion \cite{MoserImpactofAlignmentonSPAAM:2108}. 
The alignment setup can be classified as environment-centric and user-centric. Under the environment-centric setup, target points are fixed at designed world locations \cite{MoserImpactofAlignmentonSPAAM:2108}. Users have to adjust their line-of-sight or on-screen reticles for alignment, thereby moving a lot between points. In contrast, under the user-centric setup, virtual cursors are displayed on the screen relative to users. Users can move a handheld target in the peripersonal space for alignment while seated \cite{o2013user, MoserImpactofAlignmentonSPAAM:2108}, thereby reducing the postural sway and physical exertion. Moser et al. suggested that the user-centric alignment setup can yield significantly more accurate and consistent results \cite{MoserImpactofAlignmentonSPAAM:2108} the environment-centric setup. 

Much research effort has been devoted to improving the user-centric alignment.
Grubert et al. introduced a Multi-Point Alignment Method (MPAAM): users simultaneously align an object to multiple landmarks, so that the same number of correspondences can be collected with fewer alignments. Azimi et al. proposed a modified 3D-version of MPAAM for commercial HMDs, where the five corners of a tracked cube were to be aligned to their virtual counterparts each time \cite{azimi2017alignment}. These methods, however, often come at the price of an increased mental workload and reduced alignment quality \cite{MoserImpactofAlignmentonSPAAM:2108} mostly due to the degraded perception of relative depths when the user interacts with the AR scene \cite{CutoloHumanPnP:2015}. To remove the constraints by extra devices and be more flexible in usage, some methods exploit the user's finger as the calibration target, instead of a specially designed tool. In the works by Jun et al. \cite{jun2016calibration} and Moser et al. \cite{MoserLeapMotion:2016}, the user was asked to align his/her fingertip(s) with the displayed reticle(s). The fingertip was computed from the raw samplings of a HMD-embedded depth camera \cite{jun2016calibration} or directly obtained from a Leap Motion controller \cite{MoserLeapMotion:2016}. As one correspondence was collected per alignment, the two works required at least 8 \cite{jun2016calibration} and 25 alignments \cite{MoserLeapMotion:2016} separately for reliable calibration. Compared to the stylus alignment, fingertip alignment results in a higher calibration error, since the finger tracking is less precise and robust \cite{MoserLeapMotion:2016}.

\subsection{Homography-corrected off-axis eye-display model}
\begin{figure}[htb]
 \centering 
 \includegraphics[width=0.8\columnwidth]{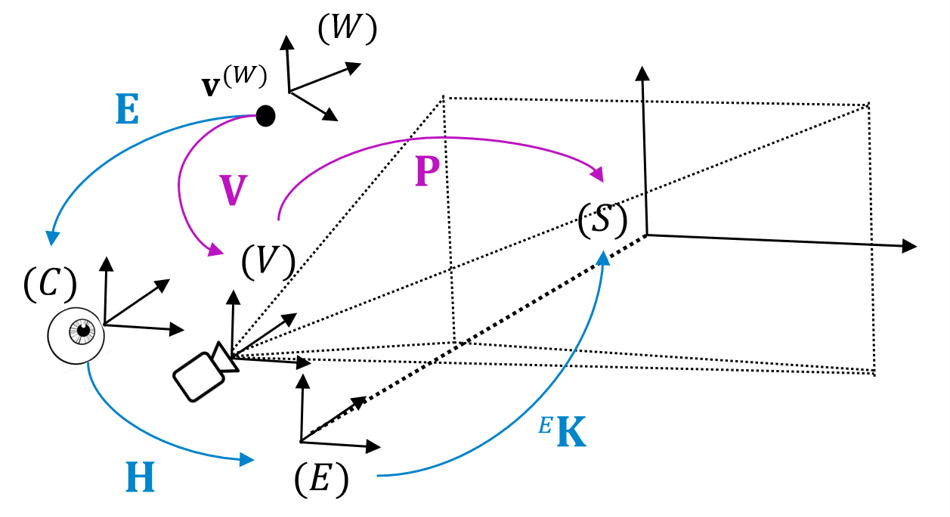}
 \caption{System overview of the OST calibration based on homography corrections. $W$: static world system; $V$: off-axis rendering camera system; $S$: display coordinate; $C$: viewpoint-display off-axis camera system; $E$: a presumed ideal on-axis camera system.}
 \label{fig:offline}
\end{figure}

\autoref{fig:offline} shows an overview of the involved coordinates in OST calibration. The viewpoint-display system $C$, which is regarded as an off-axis pinhole camera, should be calibrated first to define the correspondence between a real point $\mathbf{v}^{\scriptscriptstyle {(W)}}$ and its aligned virtual counterpart $\mathbf{m}^{\scriptscriptstyle {(S)}}$. The projection of the rendering camera $V$ can then be corrected for consistent display.     

The viewpoint can be, of course, the actual nodal point of the user's eye, or the optical centre of an eye-replacement camera. The eye-replacement camera is usually used in offline calibration to compute the hardware-related projection parameters at an initial generic viewpoint \cite{OwenDRC:2004, cutolo2020off}. In a recent work, homography correction was exploited to produce the off-axis viewpoint-display system ($C$) from an ideal on-axis camera-display system ($E$) \cite{hu2020alignment, cutolo2020off}. The projected pixels on the see-through display $S$ (i.e.,  $\mathbf{m}^{\scriptscriptstyle {(S)}}$) and the camera $C$'s frames (i.e., $\mathbf{m}^{\scriptscriptstyle {(C)}}$) are related by an homography-based warping:
\begin{equation} 
\zeta\mathbf{m}^{\scriptscriptstyle {(S)}} =\zeta\mathbf{m}^{\scriptscriptstyle {(E)}}= \underbrace{\prescript{\scriptscriptstyle {E}}{}{\mathbf{K}}~ (\mathbf{I}_{\scriptscriptstyle {3\times3}} + \frac{\prescript{\scriptscriptstyle E}{\scriptscriptstyle C}{\mathbf{t}}(\mathbf{n}_{\scriptscriptstyle S})^{\texttt T}}{d_{\scriptscriptstyle{CS}}}) 
\prescript{\scriptscriptstyle E}{\scriptscriptstyle C}{\mathbf{R}}~ {(\prescript{\scriptscriptstyle {C}}{}{\mathbf{K}})}^{\scriptscriptstyle {-1}}}_{\prescript{\scriptscriptstyle {E}}{\scriptscriptstyle C}{\mathbf{H}}}~\underbrace{ \prescript{\scriptscriptstyle {C}}{}{\mathbf{K}}~
\begin{bmatrix}
\prescript{\scriptscriptstyle C}{\scriptscriptstyle W}{\mathbf{R}}  & \prescript{\scriptscriptstyle C}{\scriptscriptstyle W}{\mathbf{t}} 
\end{bmatrix}
\mathbf{v}^{\scriptscriptstyle {(W)}}}_{\mathbf{m}^{\scriptscriptstyle {(C)}}}
\end{equation}
where $\mathbf{n}_{\scriptscriptstyle S}$ is the normal unit vector of the display in $E$ reference system (i.e., $(0,0,1$), $d_{\scriptscriptstyle{CS}}$ is the distance between $C$ and $S$. $\prescript{\scriptscriptstyle {E}}{}{\mathbf{K}}$ is the on-axis intrinsic and $\prescript{\scriptscriptstyle {C}}{}{\mathbf{K}}$ is the off-axis intrinsic. $\prescript{\scriptscriptstyle {E}}{\scriptscriptstyle C}{\mathbf{H}}$ is the planar homography between $C$ and $E$. 

Reorganising the above equation we have:
\begin{equation} 
\zeta\mathbf{m}^{\scriptscriptstyle {(S)}} = \prescript{\scriptscriptstyle {E}}{}{\mathbf{K}}~ \mathbf{H}~ \mathbf{E}~ \mathbf{v}^{\scriptscriptstyle {(W)}}
\end{equation}
where $\mathbf{H}$ is a matrix associated with homography correction. Following the definition of axis direction in \autoref{fig:offline}:
\begin{equation} 
\mathbf{H} = \mathbf{I}_{\scriptscriptstyle {3\times3}} + \frac{\prescript{\scriptscriptstyle E}{\scriptscriptstyle C}{\mathbf{t}}(\mathbf{n}_{\scriptscriptstyle E})^{\texttt T}}{d_{\scriptscriptstyle{CS}}}
=
\begin{bmatrix}
\frac{z_{\scriptscriptstyle{CS}}}{z_{ ES}} & 0 & -\frac{x_{CE}}{z_{ ES}}  \\
0   & \frac{z_{ CS}}{z_{ ES}}   & -\frac{y_{CE}}{z_{ ES}}   \\
0   & 0     & 1
\end{bmatrix}	
\label{eq:H}
\end{equation}
$x_{\scriptscriptstyle{CE}},y_{\scriptscriptstyle{CE}}$ are the x, y components of $\mathbf{t}_{\scriptscriptstyle{CE}}^{\scriptscriptstyle{(S)}}$. $z_{\scriptscriptstyle{ES}}$ and $z_{\scriptscriptstyle{CS}}$ are the z components of $\mathbf{t}_{\scriptscriptstyle{ES}}^{\scriptscriptstyle{(S)}}$ and $\mathbf{t}_{\scriptscriptstyle{CS}}^{\scriptscriptstyle{(S)}}$. 
% By their nature, $\mathbf{t}_{\scriptscriptstyle{CE}}^{\scriptscriptstyle{(S)}} + \mathbf{t}_{\scriptscriptstyle{ES}}^{\scriptscriptstyle{(S)}} = \mathbf{t}_{\scriptscriptstyle{CS}}^{\scriptscriptstyle{(S)}}$. 
$\mathbf{H}$ encapsulates the shift and scaling due to viewpoint movement, and relates the ideal $\prescript{\scriptscriptstyle {E}}{}{\mathbf{K}}$ (i.e., a constant matrix decided by hardware properties) to the actual $\prescript{\scriptscriptstyle {C}}{}{\mathbf{K}}$ dependent on the specific viewpoint. 

$\mathbf{E}$ can be expressed as: 
\begin{equation} 
\mathbf{E} =\prescript{\scriptscriptstyle E}{\scriptscriptstyle C}{\mathbf{R}}~\begin{bmatrix}
\prescript{\scriptscriptstyle C}{\scriptscriptstyle W}{\mathbf{R}}  & \prescript{\scriptscriptstyle C}{\scriptscriptstyle W}{\mathbf{t}} 
\end{bmatrix} = 
\begin{bmatrix}
\prescript{\scriptscriptstyle E}{\scriptscriptstyle W}{\mathbf{R}}  & \prescript{\scriptscriptstyle E}{\scriptscriptstyle C}{\mathbf{R}}\prescript{\scriptscriptstyle C}{\scriptscriptstyle W}{\mathbf{t}} 
\end{bmatrix}
\label{eq:E}
\end{equation}
Notably, the rotational contribution of $\mathbf{E}$ (i.e., $\prescript{\scriptscriptstyle E}{\scriptscriptstyle W}{\mathbf{R}}$) is independent on the rotation of $\mathbf{C}$.
Furthermore, if $\mathbf{C}$ presents an actual user's eye rather than a camera, $\prescript{\scriptscriptstyle E}{\scriptscriptstyle C}{\mathbf{R}}=\mathbf{I}_{\scriptscriptstyle {3\times3}}$ and thus $\mathbf{E}$ is equivalent to the extrinsic matrix of $\mathbf{C}$ in the world.
% , and it also accounts for the deviations of the real optical features
% of the see-through display from the ones dictated by the display manufacturer's specifics.
%%%%%%%%%%%%%%%%%%%

% Such a model can be used for the offline OST calibration at an initial arbitrary viewpoint (i.e., the optical centre of $C$). 

\section{Material and methods}
\subsection{Automatic hand tracking and cursor generation}
\begin{figure}[htb]
 \centering 
 \includegraphics[width=\columnwidth]{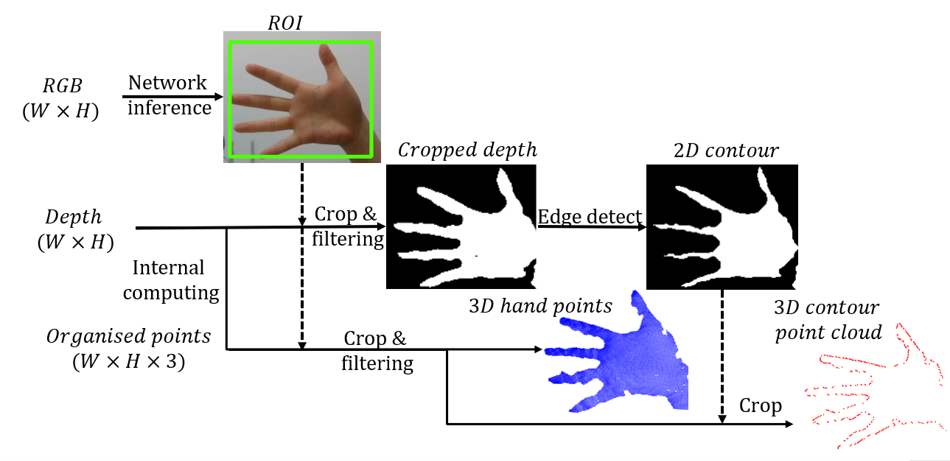}
 \caption{The segmentation of 3D hand points and contour points from RGBD camera captures.}
 \label{fig:hand}
\end{figure}

Instead of relying on a specially-made calibration tool, we exploit the user’s bare-hand sampled by an RGBD camera $D$ for the online alignment collection. A MobileNetV1-based SSD network \cite{liu2016ssd} was trained on the Oxford Hands Dataset \cite{mittal2011hand} to locate the region of interest (ROI) for the hand in the captured RGB frames. To improve the robustness of our ROI predictor, images with different background, gestures and skin tones were trained with augmented hue, saturation, brightness and contrast. The network achieves a mean average precision (mAP) of 0.958 with 0.5 intersection over union (IOU) on the test dataset.

The hand points segmentation procedure is shown in \autoref{fig:hand}. The RGB frames in a shape of height(H)$\times$width(W) was first fed to the trained network to detect a rectangular ROI. 
The depth frames (H$\times$W$\times$3) were cropped by the predicted ROI and filtered by the 0.35-0.7m depth threshold (i.e., the intersection between arm reach distance and depth camera working range), resulting in the tracked 3D hand points. Such a tracking procedure requires users to raise their bare hand in the mid-air, so that no other object exists in the ROI cropping box within the threshold distance. 

Instead of displaying a hand cursor with a pre-designed geometry at fixed locations, we allow the user to generate customised cursor with their own hand at self-dedicated locations. To this aim, the cropped and filtered depth frames were converted to binary masks and processed by a contour-finding algorithm \cite{suzuki1985topological}. The cropped depth frames were further segmented by the detected contour mask, resulting in a number of 3D contour points that could be displayed as a reference cursor by the HMD. 
\begin{figure*}[htbp]
\centering
\subfigure[]{\includegraphics[height=3cm]{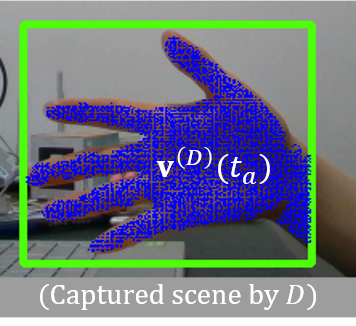}}
\subfigure[]{\includegraphics[height=3cm]{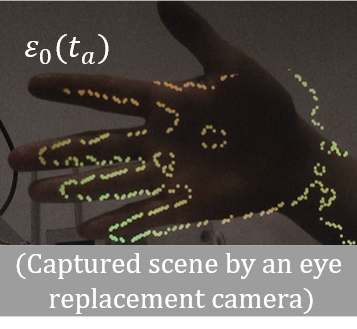}}
\subfigure[]{\includegraphics[height=3cm]{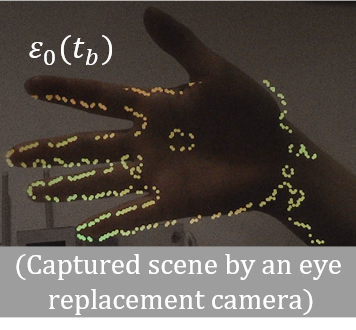}}
\subfigure[]{\includegraphics[height=3cm]{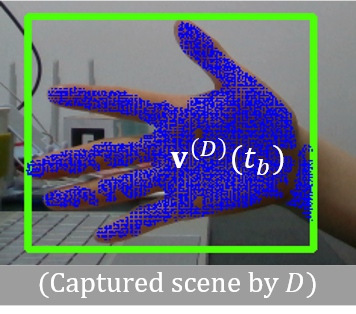}}
\caption{The online correspondence collection by alignment with the user's hand. An eye-replacement camera recorded the ``user'' perspective. (a) The display of the ``hand cursor'' using the tracked hand at frame $t_{\scriptscriptstyle a}$. (b) When the display is yet to be calibrated, the user will perceive a misalignment between hand and cursor. (c) The user moves his/her hand so that it appears as aligned with the cursor. (d) When alignment is achieved at $t_{\scriptscriptstyle b}$, the 3D hand points are taken for point-to-point registration.}
\label{fig:display}
\end{figure*}

\subsection{Update of the homography-corrected off-axis model}
Our online calibration update contains two steps: optimising the viewpoint shift-related parameters according to the homography-corrected off-axis eye-display model, and updating the offline calibrated projection by the solved parameters. We will first introduce our model for the viewpoint shift-induced display update, and then give the details of two calibration steps. 

Starting from a camera-based offline calibration as in \cite{hu2020alignment, OwenDRC:2004}, the matrix $\mathbf{H}$ and $\mathbf{E}$ can be individually updated by a matrix $\mathbf{U}$ and $\mathbf{Q}$ that both encapsulate the viewpoint shift $ [x_{C_{\scriptscriptstyle 0}C_{\scriptscriptstyle 1}}, y_{ C_{\scriptscriptstyle 0}C_{\scriptscriptstyle 1}}, z_{ C_{\scriptscriptstyle 0}C_{\scriptscriptstyle 1}}]^{\scriptscriptstyle T}$ from the offline-calibrated viewpoint $C_{\scriptscriptstyle 0}$ to the new viewpoint $C_{\scriptscriptstyle1}$ (i.e., the user's eye actual position): we define the update matrix $\mathbf{U}$ for $\mathbf{H}$, so that $\mathbf{H}_{\scriptscriptstyle 1} =
\mathbf{H}_{\scriptscriptstyle 0} \mathbf{U}$. According to \autoref{eq:H}, $\mathbf{U}$ can be expressed as (in 4$\times$4 OpenGL convention):
\begin{align} 
\mathbf{U} =
{\mathbf{H}_{\scriptscriptstyle 0}}^{\scriptscriptstyle {-1}}~\mathbf{\large H}_{\scriptscriptstyle 1} &=
\begin{bmatrix}
1-\frac{z_{ C_{\scriptscriptstyle 0}C_{\scriptscriptstyle 1}}}{z_{ C_{\scriptscriptstyle 0}S}} & 0 & \frac{x_{ C_{\scriptscriptstyle 0}C_{\scriptscriptstyle 1}}}{z_{ C_{\scriptscriptstyle 0}S}}  & 0 \\
0   & 1-\frac{z_{ C_{\scriptscriptstyle 0}C_{\scriptscriptstyle 1}}}{z_{ C_{\scriptscriptstyle 0}S}} & \frac{y_{ C_{\scriptscriptstyle 0}C_{\scriptscriptstyle 1}}}{z_{ C_{\scriptscriptstyle 0}S}}  & 0 \\
0   & 0     & 1  & 0  \\
0   & 0     & 0  & 1  \\
\end{bmatrix}
% \\
% &=
% \begin{bmatrix}
% 1-\phi_3\phi_4 & 0 & \phi_1\phi_4  & 0 \\
% 0   & 1-\phi_3\phi_4 & \phi_2\phi_4  & 0 \\
% 0   & 0     & 1  & 0  \\
% 0   & 0     & 0  & 1  \\
% \end{bmatrix}	
% \label{eq:U}
\end{align}
where $z_{C_{\scriptscriptstyle 0}S}$ is the distance from the offline calibration camera to the virtual display (i.e., the display focal plane).

We define the update matrix $\mathbf{Q}$ for $\mathbf{E}$, so that 
$\mathbf{E}_{\scriptscriptstyle 1} = \mathbf{Q}
\mathbf{E}_{\scriptscriptstyle 0}
$. According to \autoref{eq:E}:
\begin{align} 
\mathbf{Q} = \mathbf{E}_{\scriptscriptstyle 1}{\mathbf{E}_{\scriptscriptstyle 0}}^{\scriptscriptstyle{-1}}
= 
\begin{bmatrix}
1   & 0     & 0     & -x_{C_{\scriptscriptstyle 0}C_{\scriptscriptstyle 1}} \\
0   & 1     & 0     & -y_{C_{\scriptscriptstyle 0}C_{\scriptscriptstyle 1}} \\
0   & 0     & 1     & -z_{C_{\scriptscriptstyle 0}C_{\scriptscriptstyle 1}}  \\
0   & 0     & 0     & 1  \\
\end{bmatrix}
% \\
% &=
% \begin{bmatrix}
% 1 & 0 & 0  & -\phi_1 \\
% 0   & 1 & 0  & -\phi_2 \\
% 0   & 0     & 1  & -\phi_3  \\
% 0   & 0     & 0  & 1  \\
% \end{bmatrix}
\label{eq:Q}
\end{align}

Thus, at a frame $t$, the virtual pixel $\mathbf{m}_{\scriptscriptstyle 1}$ that coherently aligns with its real counterpart from the new perspective is:
\begin{equation}
\zeta\mathbf{m}_{\scriptscriptstyle 1}(t) =\prescript{\mathrm{on-E}}{}{\mathbf{K}} \mathbf{H}_{\scriptscriptstyle 1} ~
\mathbf{E}_{\scriptscriptstyle 1}(t)~ \mathbf{v}_{\scriptscriptstyle 1}^{\scriptscriptstyle {(W)}}(t)= \prescript{\mathrm{on-E}}{}{\mathbf{K}} \mathbf{H}_{\scriptscriptstyle 0}~ \mathbf{U}\mathbf{Q}~ \mathbf{E}_{\scriptscriptstyle 0}(t) ~ \mathbf{v}_{\scriptscriptstyle 1}^{\scriptscriptstyle {(W)}}(t)
\label{eq:pipeline}
\end{equation}

Let $\phi_1 = x_{ C_{\scriptscriptstyle 0}C_{\scriptscriptstyle 1}}, \phi_2 = y_{ C_{\scriptscriptstyle 0}C_{\scriptscriptstyle 1}}, \phi_3 = z_{ C_{\scriptscriptstyle 0}C_{\scriptscriptstyle 1}}$ and $\tilde{\phi_4} = 1 / z_{C_{\scriptscriptstyle 0}S}$:
\begin{align} 
\mathbf{UQ} = 
\begin{bmatrix}
1-\phi_3\tilde{\phi_4} & 0 & \phi_1\tilde{\phi_4}  & -\phi_1 \\
0   & 1-\phi_3\tilde{\phi_4} & \phi_2\tilde{\phi_4}  & -\phi_2 \\
0   & 0     & 1  & -\phi_3  \\
0   & 0     & 0  & 1  \\
\end{bmatrix}
\label{eq:UQ}
\end{align}
where $\tilde{\phi_4}$ can be calibrated offline. $[\phi_1, \phi_2, \phi_3]^{\scriptscriptstyle T}$ is the unknown vector to be optimised online by user alignment. 

\subsubsection{Step 1: parameter optimisation by online alignment}
A key aspect is how to optimise $[\phi_1, \phi_2, \phi_3]^{\scriptscriptstyle T}$. Unlike some other alignment-based methods that focus on a full 3D world-2D pixel correspondence scheme (e.g., by collecting a set of 3D-2D correspondences that reflect the correct projection), we focus on the relative transformation between two 3D point clouds that are projected at the same pixel location before and after alignment correction (i.e., opting for a 3D-3D scheme). We adopt an object-wise alignment based on the tracked hand contour rather than several point-wise alignments.

Starting from the offline calibration obtained with $C_{\scriptscriptstyle 0}$, the 3D hand contour sampled at a moment $t_{\scriptscriptstyle a}$ is transformed into the world system and dynamically rendered by the visor as a cursor $\mathbf{\epsilon}_{\scriptscriptstyle 0}(t)$:
\begin{equation}
\zeta\mathbf{\epsilon}_{\scriptscriptstyle 0}(t)
=\prescript{\mathrm{on-E}}{}{\mathbf{K}}\mathbf{H}_{\scriptscriptstyle 0} \mathbf{E}_{\scriptscriptstyle 0}(t) \times \mathbf{v}^{\scriptscriptstyle {(W)}}(t_{\scriptscriptstyle a})
\label{eq:1}
\end{equation}
Since the display is not calibrated to the user's real viwpoint, a parallax-related misalignment will exist between the displayed cursor and the user-perceived hand (\autoref{fig:display} (b)). 

At a moment $t_{\scriptscriptstyle b}$, the user is asked to align the hand with the cursor generated at $t_{\scriptscriptstyle a}$ (\autoref{fig:display} (c)). The correctly updated eye-display model should give the exact virtual-to-real correspondence as collected by the user:
\begin{align}
\zeta'\mathbf{\epsilon}_{\scriptscriptstyle 0}(t_{\scriptscriptstyle b})
&=  \prescript{\mathrm{on-E}}{}{\mathbf{K}} \mathbf{H}_{\scriptscriptstyle 1} \mathbf{E}_{\scriptscriptstyle 1}(t_{\scriptscriptstyle b}) \times \mathbf{v}^{\scriptscriptstyle {(W)}}(t_{\scriptscriptstyle b})
\label{eq:2}
\end{align}
Note that the displayed pixels are uniquely determined by their Cartesian expression $\mathbf{\epsilon}_{\scriptscriptstyle 0}$. The new scaling factor $\zeta'$ (i.e., the distance between $\mathbf{v}^{\scriptscriptstyle {(W)}}(t_{\scriptscriptstyle b})$ and the principal plane for $C_{\scriptscriptstyle 1}$) can be different from $\zeta$ (i.e., the distance between $\mathbf{v}^{\scriptscriptstyle {(W)}}(t_{\scriptscriptstyle a})$ and the principal plane for $C_{\scriptscriptstyle 0}$). If the offline calibration is performed at a proper viewpoint position within the eye-box and in the eye-relief distance from the OC, the translation from $\mathbf{v}^{\scriptscriptstyle {(W)}}(t_{\scriptscriptstyle a})$ to $\mathbf{v}^{\scriptscriptstyle {(W)}}(t_{\scriptscriptstyle b})$ in depth (i.e., $\zeta'-\zeta$) will be much smaller than the distance from $\mathbf{v}^{\scriptscriptstyle {(W)}}(t_{\scriptscriptstyle a})$ to $C_{\scriptscriptstyle 1}$ (i.e., $\zeta$). Therefore, we reasonably assume that $\zeta'/\zeta \approx 1$. Combining \autoref{eq:1}, \autoref{eq:2} and \autoref{eq:pipeline}, at the alignment moment $t_{\scriptscriptstyle b}$:
\begin{equation}
 \prescript{\mathrm{on-E}}{}{\mathbf{K}} \mathbf{H}_{\scriptscriptstyle 0} \mathbf{E}_{\scriptscriptstyle 0}(t_{\scriptscriptstyle b}) ~ \mathbf{v}^{\scriptscriptstyle {(W)}}(t_{\scriptscriptstyle a})
= \prescript{\mathrm{on-E}}{}{\mathbf{K}} \mathbf{H}_{\scriptscriptstyle 0} \mathbf{U}\mathbf{Q} \mathbf{E}_{\scriptscriptstyle 0}(t_{\scriptscriptstyle b}) ~ \mathbf{v}^{\scriptscriptstyle {(W)}}(t_{\scriptscriptstyle b})
\label{eq:align}
\end{equation}

Considering the transformation between two obtained point clouds by a matrix $\mathbf{M}$: 
\begin{equation}
\mathbf{v}^{\scriptscriptstyle {(W)}}(t_{\scriptscriptstyle a}) = \mathbf{M}~ \mathbf{v}^{\scriptscriptstyle {(W)}}(t_{\scriptscriptstyle b})
\label{eq:M}
\end{equation}
According to \autoref{eq:align}:
\begin{equation}
    \mathbf{M} = \mathbf{E}_{\scriptscriptstyle 0}(t_{\scriptscriptstyle b})^{\scriptscriptstyle {-1}} ~\underbrace{
    \begin{bmatrix}
1-\phi_3\tilde{\phi_4} & 0 & \phi_1\tilde{\phi_4}  & -\phi_1 \\
0   & 1-\phi_3\tilde{\phi_4} & \phi_2\tilde{\phi_4}  & -\phi_2 \\
0   & 0     & 1  & -\phi_3  \\
0   & 0     & 0  & 1  \\
\end{bmatrix}}_{\mathbf{UQ}}
~\mathbf{E}_{\scriptscriptstyle 0}(t_{\scriptscriptstyle b}) 
\label{eq:M}
\end{equation}

A single alignment can provide a massive number of corresponding $\mathbf{v}^{\scriptscriptstyle {(W)}}(t_{\scriptscriptstyle a})$ and $  \mathbf{v}^{\scriptscriptstyle {(W)}}(t_{\scriptscriptstyle b})$ for the unknown optimisation. Optimising the three unknowns ($\phi_1, \phi_2, \phi_3$) is equivalent to solving the 3 DOF-relative transformation $\mathbf{M}$ that is in a special form given by \autoref{eq:M}. 

Since the two collected point sets are unpaired, we adopted the iterative correspondence search used by iterative closest point (ICP) registration. Let $\mathbf{p}_{\scriptscriptstyle i}=\mathbf{v}_{\scriptscriptstyle i}^{\scriptscriptstyle {(W)}}(t_{\scriptscriptstyle b})$  ($i=1,2,...N$, i.e., the number of points) and $\mathbf{X} = \mathbf{E}_{\scriptscriptstyle 0}(t_{\scriptscriptstyle b})$ for convenience. The two point clouds are first initially aligned by their mean positions. $\mathbf{q}_{\scriptscriptstyle i}$ is the nearest neighbour point of transformed $\mathbf{p}_{\scriptscriptstyle i}$ searched in $\mathbf{v}^{\scriptscriptstyle {(W)}}(t_{\scriptscriptstyle a})$ during every iteration $k$ based on the k-d tree \cite{bentley1975multidimensional}. The point-to-point registration error obtained by the current estimation $\mathbf{M}_{\scriptscriptstyle k}$ is defined as:
\begin{align}
e_{\scriptscriptstyle k} 
& = \sum_{i=1}^{N} {\norm{\mathbf{M}_{\scriptscriptstyle k} \mathbf{p}_{\scriptscriptstyle i} - \mathbf{q}_{\scriptscriptstyle i}}}^{\scriptscriptstyle 2} = \sum_{i=1}^{N} (\mathbf{M}_{\scriptscriptstyle k} \mathbf{p}_{\scriptscriptstyle i} - \mathbf{q}_{\scriptscriptstyle i})\cdot (\mathbf{M}_{\scriptscriptstyle k} \mathbf{p}_{\scriptscriptstyle i}- \mathbf{q}_{\scriptscriptstyle i}) \\
& \approx  \sum_{i=1}^{N} (\mathbf{M}_{\scriptscriptstyle k} \mathbf{p}_{\scriptscriptstyle i} - \mathbf{q}_{\scriptscriptstyle i})\cdot (\mathbf{M}_{\scriptscriptstyle {k-1}} \mathbf{p}_{\scriptscriptstyle i} - \mathbf{q}_{\scriptscriptstyle i})
\end{align}
We update one part of $e_{\scriptscriptstyle k}$ and approximate the other part with the latest result $\mathbf{M}_{\scriptscriptstyle {k-1}}$, to make sure $e_{\scriptscriptstyle k}$ is linear with $[\phi_1, \phi_2, \phi_3]^{\scriptscriptstyle T}$ (otherwise non-linear optimisation is computational expensive for numerous points).
$\mathbf{\lambda}_{\scriptscriptstyle i} = (\mathbf{M}_{\scriptscriptstyle {k-1}}\mathbf{p}_{\scriptscriptstyle i} - \mathbf{q}_{\scriptscriptstyle i})$ is a constant vector in the current iteration $k$. Therefore:
\begin{align}
e_{\scriptscriptstyle k} = \sum _{i=1}^{N}(\mathbf{M}_{\scriptscriptstyle k} \mathbf{p}_{\scriptscriptstyle i} - \mathbf{q}_{\scriptscriptstyle i})\cdot \mathbf{\lambda}_{\scriptscriptstyle i} 
% &= \sum _{i=1}^{N}{[\mathbf{\lambda}_{\scriptscriptstyle i}^{\scriptscriptstyle T}(\mathbf{X}^{\scriptscriptstyle {-1}} \mathbf{UQ} \mathbf{X} \mathbf{p}_{\scriptscriptstyle i} - \mathbf{q}_{\scriptscriptstyle i}) ]}\\
= \sum _{i=1}^{N}{[\mathbf{\lambda}_{\scriptscriptstyle i}^{\scriptscriptstyle T}\mathbf{X}^{\scriptscriptstyle {-1}}(\mathbf{UQ} \mathbf{X} \mathbf{p}_{\scriptscriptstyle i} - \mathbf{X}\mathbf{q}_{\scriptscriptstyle i})]}
\end{align}
Let $\mathbf{\lambda}_{\scriptscriptstyle i}^{\scriptscriptstyle T}\mathbf{X}^{\scriptscriptstyle {-1}} = \mathbf{\lambda '}_{\scriptscriptstyle i}^{\scriptscriptstyle T}, \mathbf{s}_{\scriptscriptstyle i} = \mathbf{X} \mathbf{p}_{\scriptscriptstyle i}$ and $\mathbf{d}_{\scriptscriptstyle i} = \mathbf{X} \mathbf{q}_{\scriptscriptstyle i}$,
\begin{align}
e_{\scriptscriptstyle k}= \sum _{i=1}^{N}\mathbf{\lambda '}_{\scriptscriptstyle i}^{\scriptscriptstyle T}(\mathbf{UQ} \mathbf{s}_{\scriptscriptstyle i} - \mathbf{d}_{\scriptscriptstyle i})= \sum _{i=1}^{N}[\mathbf{A} \mathbf{x} - \mathbf{b}]
\end{align}

where
\begin{equation}
\mathbf{A}
= -\begin{bmatrix}
\dots&\dots&\dots\\
\mathbf{\lambda'}_{\scriptscriptstyle {jx}}(1-\tilde{\phi}_4\mathbf{s}_{\scriptscriptstyle {jz}})&    \mathbf{\lambda'}_{\scriptscriptstyle {jy}}(1-\tilde{\phi}_4\mathbf{s}_{\scriptscriptstyle {jz}})&    \tilde{\phi}_4(\mathbf{\lambda'}_{\scriptscriptstyle {jx}}\mathbf{s}_{\scriptscriptstyle {jx}}
+\mathbf{\lambda'}_{\scriptscriptstyle {jy}}\mathbf{s}_{\scriptscriptstyle {jy}})+\mathbf{\lambda'}_{\scriptscriptstyle {jz}}\\
\dots&\dots&\dots
\end{bmatrix}
\end{equation}

\begin{equation}
\mathbf{x}
= \begin{bmatrix}
\phi_1\\\phi_2\\\phi_3
\end{bmatrix}
~
\mathbf{b}
= \begin{bmatrix}
\dots\\
(\mathbf{d}_{\scriptscriptstyle {j}}-\mathbf{s}_{\scriptscriptstyle {j}})\cdot\mathbf{\lambda'}_{\scriptscriptstyle {j}}\\
\dots
\end{bmatrix}
\end{equation}
$e_{\scriptscriptstyle k}$ can thus be minimised through a standard least-square problem $\mathbf{A} \mathbf{x} = \mathbf{b}$ by Single Value Decomposition (SVD). The iteration will terminate if $1>\frac{e_{\scriptscriptstyle k}}{e_{\scriptscriptstyle {k-1}}}>0.9999$, or the maximum iteration limit (e.g., 800) has been reached.

Note that the special expression of $\mathbf{M}$ indicates that the viewpoint shift $[\phi_1, \phi_2, \phi_3]^{\scriptscriptstyle T}$ hardly causes a relative rotation in the tracked 3D space: as $\phi_1, \phi_2, \phi_3$ are usually up to 2 cm due to the limited size of the eye box, and the display focal length is in the meter range (e.g., $z_{C_{\scriptscriptstyle 0}S}\approx$ 2 m for Microsoft HoloLens), $\phi_1\tilde{\phi_4}, \phi_2\tilde{\phi_4}, \phi_3\tilde{\phi_4}$ will be tiny. Thus, the first 3$\times$3 part of $\mathbf{UQ}$ approximates an identity matrix, resulting in a nearly identity rotation in $\mathbf{M}$. 
In other words, only if the user's hand is mainly translated rather than rotated from the initial cursor generation pose $\mathbf{v}^{\scriptscriptstyle {(W)}}(t_{\scriptscriptstyle a})$ to the alignment pose $\mathbf{v}^{\scriptscriptstyle {(W)}}(t_{\scriptscriptstyle b})$, accurate alignment can be achieved. We thus call such registration-based optimisation as ``rotation-constrained ICP'' (rcICP) method. We will prove later, through simulation, that such a method can yield robust estimates for $\phi_1, \phi_2, \phi_3$ even when some undesired relative mis-rotations exist between $\mathbf{v}^{\scriptscriptstyle {(W)}}(t_{\scriptscriptstyle a})$ and $\mathbf{v}^{\scriptscriptstyle {(W)}}(t_{\scriptscriptstyle b})$.
\subsubsection{Step 2: update of OST projection matrix}
After solving the unknowns, the $\mathbf{UQ}$ matrix can be computed by \autoref{eq:UQ}. The projection parameters of the HMD rendering camera can be updated by the equilibrium between the blue and purple transformation lines as shown in \autoref{fig:offline}:
\begin{equation} 
\prescript{\mathrm{on-E}}{}{\mathbf{K}} \mathbf{H}_{\scriptscriptstyle 1} ~
\mathbf{E}_{\scriptscriptstyle 1}(t)~ \mathbf{v}^{\scriptscriptstyle {(W)}}(t)
= \mathbf{P}_{\scriptscriptstyle 1}~ \prescript{\scriptscriptstyle {V}}{\scriptscriptstyle W}{\mathbf{T}}(t) ~ \mathbf{v}^{\scriptscriptstyle {(W)}}(t)
\end{equation}
\begin{equation}
  \mathbf{P}_{\scriptscriptstyle 1} =
\prescript{\mathrm{on-E}}{}{\mathbf{K}} \mathbf{H}_{\scriptscriptstyle 0} \mathbf{UQ}
\mathbf{E}_{\scriptscriptstyle 0}(t) {(\prescript{\scriptscriptstyle {V}}{\scriptscriptstyle W}{\mathbf{T}}(t))} ^{\scriptscriptstyle {-1}}
=  \prescript{\mathrm{on-E}}{}{\mathbf{K}} \mathbf{H}_{\scriptscriptstyle 0}\mathbf{UQ}
\begin{bmatrix}
\mathbf{I}_{\scriptscriptstyle {3\times3}} &\mathbf{t}_{\scriptscriptstyle {C_{\scriptscriptstyle 0}V}}^{\scriptscriptstyle {(S)}}\\
0&1
\end{bmatrix}
\label{eq:proj}
\end{equation}
where $\mathbf{t}_{\scriptscriptstyle {C_{\scriptscriptstyle 0}V}}^{\scriptscriptstyle {(S)}}$ is the translation between the initial viewpoint $C_{\scriptscriptstyle 0}$ and rendering camera $V$, which is invariant to the online device and target tracking (i.e., $t$-independent), and is computed by the offline calibration.
%%%%%%%%%%%%%%%%%%%%%%%%%%%%%%%%%%
\section{Simulation test on rotation-constrained ICP}
A simulation test was carried out to evaluate the performance of rcICP in recovering the ground truth (gt) value of $[\phi_1, \phi_2, \phi_3]^{\scriptscriptstyle T}$ when relative rotation exists between collected $\mathbf{v}^{\scriptscriptstyle {(W)}}(t_{\scriptscriptstyle a})$ and $\mathbf{v}^{\scriptscriptstyle {(W)}}(t_{\scriptscriptstyle b})$ due to user misalignment. An automatically segmented hand point cloud was taken as the source $\mathbf{v}^{\scriptscriptstyle {(W)}}(t_{\scriptscriptstyle b})$. A matrix $\mathbf{UQ}_{\scriptscriptstyle {gt}}$ was designed with arbitrary known values of $[\phi_1, \phi_2, \phi_3]^{\scriptscriptstyle T}_{\scriptscriptstyle {gt}}$. $\mathbf{X}$ was taken as the extrinsic transformation $\mathbf{E}(t_{\scriptscriptstyle b})$ when the hand cloud was captured. After transforming source points by the designed $\mathbf{X}^{\scriptscriptstyle{-1}}\mathbf{UQ}_{\scriptscriptstyle {gt}}\mathbf{X}$, the points were additionally rotated around an arbitrary point in the transformed cloud and about a random 3D axis to simulate the imperfect user alignment that contains certain relative rotation. The final transformed points were denoted as target points $\mathbf{v}^{\scriptscriptstyle {(W)}}(t_{\scriptscriptstyle a})$. $\mathbf{v}^{\scriptscriptstyle {(W)}}(t_{\scriptscriptstyle b})$ and $\mathbf{v}^{\scriptscriptstyle {(W)}}(t_{\scriptscriptstyle a})$ were registered by the proposed rcICP method.

For comparison, the traditional ICP method was also investigated. The registration gives a transformation matrix $\mathbf{M}_{\scriptscriptstyle {ICP}}$. The equivalent $\mathbf{UQ}_{\scriptscriptstyle{ICP}} = \mathbf{X}\mathbf{M}_{\scriptscriptstyle {ICP}}\mathbf{X}^{\scriptscriptstyle{-1}}$. $\phi_1, \phi_2, \phi_3$ were taken as the translational components of the $\mathbf{UQ}_{\scriptscriptstyle{ICP}}$ matrix. The optimisation error was computed as the absolute difference between the optimised $\phi_1, \phi_2, \phi_3$ and their reference gt values.

\begin{figure}[htb]
 \centering 
 \includegraphics[height=4.8cm, width=\columnwidth]{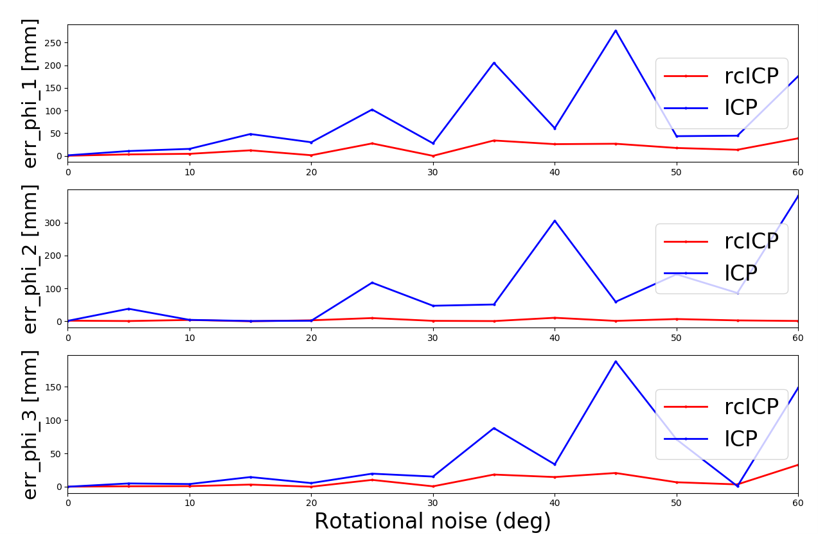}
 \caption{The effect of rotational noises on obtained $\phi_1, \phi_2$ and $\phi_3$. Data were taken with 5$^{\circ}$ incremental. }
 \label{fig:noise}
\end{figure}

\begin{figure}[htb]
 \centering 
 \includegraphics[width=\columnwidth]{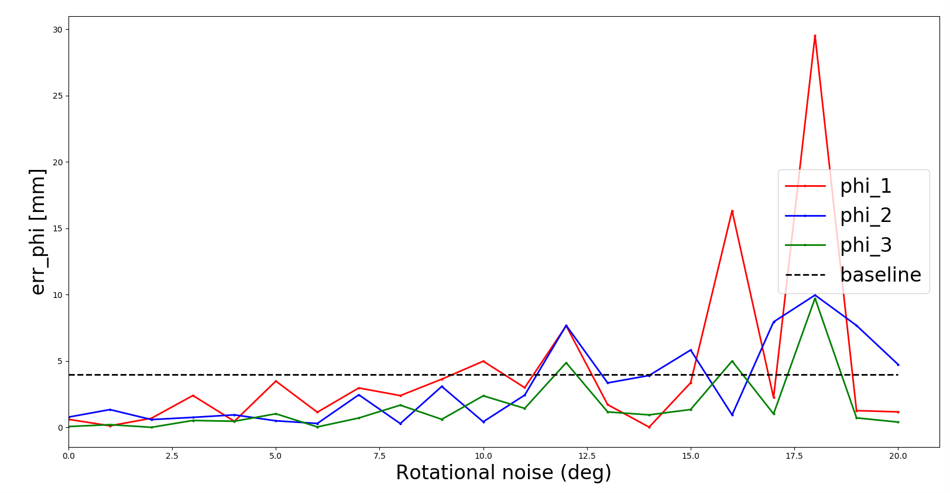}
 \caption{The accuracy of rcICP-based optimisation with up to 20$^{\circ}$ rotational noises, compared to a 4 mm baseline. Data were taken with 1$^{\circ}$ incremental. }
 \label{fig:noise2}
\end{figure}

\autoref{fig:noise} shows the result obtained under rotational noise of different magnitudes ranging from 0 to 60$^\circ$ and a random distribution about the x, y and z axes. As expected, compared to ICP, the proposed rcICP method is less sensitive to the rotational misalignment between source and target points. \autoref{fig:noise2} shows a more detailed investigation of the rcICP-based optimisation. The result indicates that our method can achieve a 4 mm accuracy (i.e., a baseline chosen as the literature-level error of alignment-based eye position estimation \cite{MoserImpactofAlignmentonSPAAM:2108}) with up to 9$^{\circ}$ rotational disturbance.


%%%%%%%%%%%%%%%%%%%%%%%%%%%%%%%%%%
\section{Implementation}
A consumer-level Intel Realsense D415 RGBD camera ($D$) was used for real-time hand detection. It can provide high-resolution depth maps up to 1280 $\times$ 720 pixels at 30 frames per second.
The camera $D$ was rigidly anchored on a commercial mixed reality headset, HoloLens generation 1, used for the AR rendering. The stereo-pose between the depth camera $D$ and the HoloLens RGB world facing camera $H$ (i.e., $ \prescript{\scriptscriptstyle {H}}{\scriptscriptstyle D}{\mathbf{T}}$) was computed offline using standard stereo-camera calibration by OpenCV \cite{opencv}. The overall re-projection error obtained by the calibrated stereo-pose was 0.52 pixels. 

The suggested offline calibration in \cite{hu2020alignment} was carried out with a manual focus eye-replacement camera (ELP 8MP webcam, Ailipu Technology Co., Ltd) placed at an arbitrary viewpoint position within the left display eye-box. The camera, hereafter referred to as the viewpoint camera, has a 1.28 arcmin/pixel angular resolution. After focusing the camera $C_{\scriptscriptstyle 0}$ on the HMD virtual display, $z_{\scriptscriptstyle{C_{\scriptscriptstyle 0}S}}$ was measured as the focal length of $C_{\scriptscriptstyle 0}$, and $\tilde{\phi_4} = 1/z_{\scriptscriptstyle{C_{\scriptscriptstyle 0}S}}$.

The real-time RGBD data stream was processed on a PC (IntelR©CoreTMi5-8250U processor with 8 giga-bytes memory and no dedicated graphics processing unit) by an online Python application. The overall hand cloud segmentation refreshes at a rate of around 22 frames per second. The recorded positions of hand contours were transformed into a static world frame $W$ that was initialised when the HoloLens application was launched. The tracked hand contour was transformed from $D$ to $W$ by:
\begin{equation}
    \mathbf{v}^{\scriptscriptstyle {(W)}}(t_{\scriptscriptstyle a})  = \prescript{\scriptscriptstyle {W}}{\scriptscriptstyle H}{\mathbf{T}}(t_{\scriptscriptstyle a}) ~ \prescript{\scriptscriptstyle {H}}{\scriptscriptstyle D}{\mathbf{T}} ~ \mathbf{v}^{\scriptscriptstyle {(D)}}(t_{\scriptscriptstyle a}) 
    \label{eq:DW}
\end{equation}
The HoloLens is capable of self-tracking in $W$ by a simultaneous localisation and mapping (SLAM)-based indoor mapping algorithm \cite{hubner2020evaluation}. The device pose $\prescript{\scriptscriptstyle {W}}{\scriptscriptstyle H}{\mathbf{T}}$
can be acquired from the HoloLens application programming interface (API). 

\begin{figure}[htb]
 \centering 
 \includegraphics[width=\columnwidth]{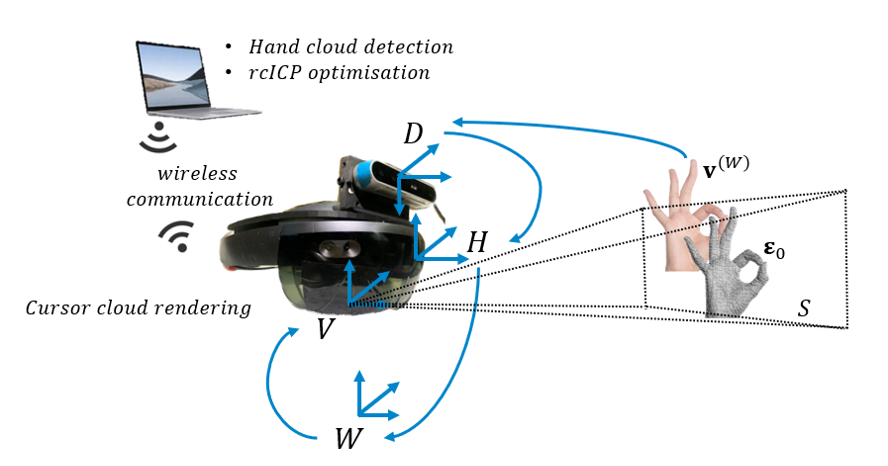}
 \caption{The system arrangement for online OST calibration with bare-hand tracking.}
 \label{fig:system}
\end{figure}

After transformation, the cloud $\mathbf{v}^{\scriptscriptstyle {(W)}}(t_{\scriptscriptstyle a})$ was rendered by the visor as a world-locked 3D cursor (\autoref{eq:1}). An HMD application was developed on the universal windows platform in Unity3D (Unity Technologies Inc.). A shader was customised for dense vertices rendering with user-specified point size. The PC and HMD application can communicate wirelessly by User Datagram Protocol (UDP). Once the alignment was achieved, the HMD application sent the extrinsic pose $\mathbf{E}_{\scriptscriptstyle 0}(t_{\scriptscriptstyle b})$ to the PC. The segmented two full-hand point clouds $\mathbf{v}^{\scriptscriptstyle {(W)}}(t_{\scriptscriptstyle a})$ and $\mathbf{v}^{\scriptscriptstyle {(W)}}(t_{\scriptscriptstyle b})$ were registered by rcICP on the PC, within a time less than 10 seconds. An overview of the system is shown in \autoref{fig:system}. 

As suggested by the simulation test results, the rcICP-based optimisation can tolerate up to 9$^{\circ}$ of hand rotation. A colour feedback was designed to guide the user alignment. The PC application ran an online ICP registration routine between $\mathbf{v}^{\scriptscriptstyle {(W)}}(t_{\scriptscriptstyle a})$ and $\mathbf{v}^{\scriptscriptstyle {(W)}}(t)$. If the magnitude of relative rotation was more than 9$^{\circ}$, the tracked hand point cloud was rendered in red rather than the default blue (as in \autoref{fig:display} (a) and (d)). Even if the user confirmed alignment, the system would not accept the current $\mathbf{v}^{\scriptscriptstyle {(W)}}(t)$ as $\mathbf{v}^{\scriptscriptstyle {(W)}}(t_{\scriptscriptstyle b})$ for the subsequent rcICP-based optimisation, and a prompt would alert the user that alignment should be carried out once more. 

For the stereo calibration, the user should align their hand with the generated cursor perceived via both displays. The rcICP-registration procedure was repeated independently for both eyes using $\mathbf{E}_{\scriptscriptstyle 0, left}(t_{\scriptscriptstyle b})$ and $\mathbf{E}_{\scriptscriptstyle 0, right}(t_{\scriptscriptstyle b})$ to optimise $(\phi_1,\phi_2,\phi_3)_{\scriptscriptstyle {left}}$ and $(\phi_1,\phi_2,\phi_3)_{\scriptscriptstyle {right}}$. The left and right projection matrices were respectively corrected by the optimised $(\tilde{\phi_1},\tilde{\phi_2},\tilde{\phi_3})_{\scriptscriptstyle {left}}$ and $(\tilde{\phi_1},\tilde{\phi_2},\tilde{\phi_3})_{\scriptscriptstyle {right}}$ according to \autoref{eq:proj}.

\section{Evaluation and result}
In this section, we present the assessment of the proposed calibration method with both an eye-replacement camera and actual users. While the camera-based evaluation is objective since the perceived OST augmentation can be directly recorded for analysis \cite{lee2014robust, makibuchi2016virc},
the user-based evaluation reflects a more realistic usage condition.  

%  The collected 3D points are reprojected by the estimated projection matrices, and the error is computed as the average distance between the resultant reprojected points and the collected 2D points. However, such method may lead to a subjective assessment as the ``ground truth'' could be biased by the quality of user alignment. Alternatively, an eye replacement camera $C$ can be placed within the visor's eye-box to record the OST augmentation for objective assessment . In this section, we report the calibration accuracy evaluated by both methods. 
\subsection{Assessment with eye-replacement camera}
\paragraph{Test design}
The same camera used for the offline calibration was used as the eye replacement camera $C'$. Unlike the eye-display system (\autoref{fig:offline}) whose orientation is defined by the virtual display $S$, the orientation of a camera $C'$ is defined by its own image plane and may mismatch the orientation of $S$. Therefore, for display calibration refinement, as the user alignment should achieve proper overlay in $S$, if such alignment is performed by observing the captured video of $C'$, those frames are to be warped to compensate for the orientation mismatch $\prescript{\scriptscriptstyle {S}}{\scriptscriptstyle {C'}}{\mathbf{R}}$: after calibrating the viewpoint camera's intrinsic $\mathbf{K}_{\scriptscriptstyle C'}$ \cite{Zhang2000:CameraCalib}, a virtual checkerboard with the known dimension and physical size was displayed at the centre of $S$ and captured by $C'$. $\prescript{\scriptscriptstyle {S}}{\scriptscriptstyle {C'}}{\mathbf{R}}$ was calibrated by a standard P\textit{n}P optimisation. After the distortion correction, the captured video frames were warped by a pure-rotational homography $\mathbf{H} = \mathbf{K}_{\scriptscriptstyle C'} ~ \prescript{\scriptscriptstyle S}{\scriptscriptstyle C'}{\mathbf{R}} ~ \mathbf{K}_{\scriptscriptstyle C'}^{\scriptscriptstyle{-1}}
$. The hand alignment was then performed within arm-reach distance (e.g., 0.5 m) by observing the warped real-time viewpoint camera video frames.

\begin{figure}[htb]
 \centering 
 \includegraphics[width=\columnwidth]{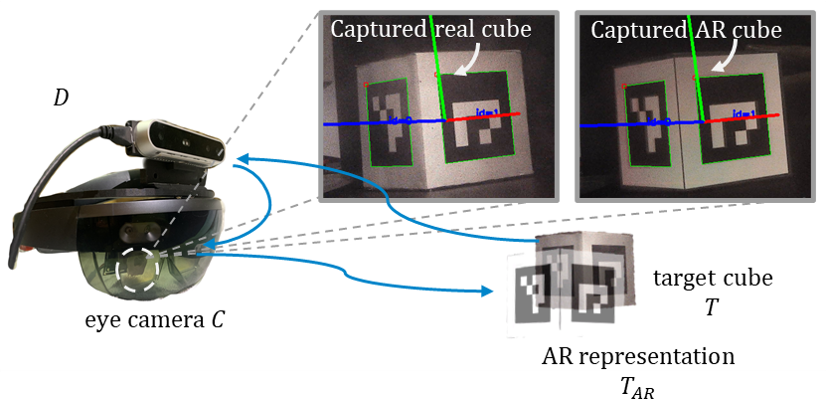}
 \caption{Evaluation of OST calibration based on the scene captured by an eye-replacement camera.}
 \label{fig:eva}
\end{figure}

As shown in \autoref{fig:eva}, a target 3D cube $T$ with an Aruco marker on every surface was tracked by $D$ in real-time. The tracked cube pose was transformed into the visor's world $W$ (\autoref{eq:DW}), according to which a virtual cube counterpart $T_{\scriptscriptstyle {AR}}$ was displayed with the calibrated projection. To avoid any interruption between the see-through scene and rendered virtual content, the viewpoint camera separately captured the real cube $T$ with the display switched-off, and the rendered virtual cube $T_{\scriptscriptstyle {AR}}$ with the real see-through background occluded. The pixel locations of marker corners were separately extracted from both images by automatic Aruco marker detection. The 3D cube orientation was then optimised from all detected corners. The rotational error $err_{\scriptscriptstyle R}$ was defined as the relative rotation between the two obtained orientations ($\prescript{T}{C'}{\mathbf{R}}$, $\prescript{T_{\scriptscriptstyle {AR}}}{C'}{\mathbf{R}}$) converted into a 3D rotation vector. We calculated the in-image misalignment of 3D translational error $err_{\scriptscriptstyle t}$ as the euclidean distance between corresponding corner pixels. It was also converted into millimetres using the pre-calibrated camera parameters of $C'$. The along-depth misalignment of $err_{\scriptscriptstyle t}$ was calculated as the absolute difference between the $z$ component of $\prescript{T}{C'}{\mathbf{t}}$ and $\prescript{T_{\scriptscriptstyle {AR}}}{C'}{\mathbf{t}}$. 

The viewpoint camera was moved in the eye-box to five different positions (i.e., centre, up, down, left and right) to cover the entire display eye-box roughly. For each $C'$ position, the proposed bare-hand alignment was redone, and the paired captures were repeated after moving and calibrating the target box to four arbitrary positions in arms-reach distance (i.e., 0.4-0.6 m away). The 20 pairs of captures obtained were processed to estimate overlay accuracy. 

\begin{table}[hbt]
\centering
\caption{Rotational and translational errors (in mm) for the calibration results in x, y and z.}
\label{tab:result}
\begin{tabular}{lllllll}
\hline
     & \multicolumn{3}{l}{$err_{\scriptscriptstyle t}$ (mm)} & \multicolumn{3}{l}{$err_{\scriptscriptstyle R}$ ($^{\circ}$)} \\ \hline
     & x          & y          & z          & x        & y       & z       \\ \hline
mean & 0.85       & 0.88       & 2.85       & 1.37     & 1.11    & 0.77    \\ \hline
std  & 0.52       & 0.53       & 2.03       & 0.95     & 0.77    & 0.76    \\ \hline
\end{tabular}
\end{table}

\begin{table*}[htb]
\centering
\caption{Comparison of 3D target misalignment with other user-centric, alignment-based OST calibration methods.}
\label{tab:compare}
\begin{tabular}{lp{4.5cm}p{3cm}p{3.5cm}p{2.7cm}}
\hline
Studies               & Collected correspondence & Alignment target & Target tracking                                                                                     & Accuracy                      \\ \hline
Ours                  & 1 $\times$ thousands of unpaired corresponding points & Contour of bare-hand  & Inside-out RGBD tracking    &  
\begin{tabular}[c]{@{}l@{}}x: 0.85$\pm$0.52 mm\\y: 0.88$\pm$0.53 mm\\z: 2.85$\pm$2.03 mm\\3D rot: 1.76$\pm$0.99$^{\circ}$\end{tabular}\\ \hline
Azimi et al. \cite{azimi2017alignment} & 4 $\times$ 5 paired corresponding points & 5 corners in a cube & Inside-out RGB tracking   & \begin{tabular}[c]{@{}l@{}}x: 0.94$\pm$0.74 mm\\y: 0.83$\pm$0.63 mm\\z: 3.51$\pm$2.67 mm \end{tabular}\\ \hline
Guo et al. \cite{GuoOSTCalibrationOnline:2019}                      & 5 $\times$ 1 paired corresponding pose of target in world and rendering space & Contour of a specially-designed box & Outside-in optical tracking                        & \begin{tabular}[c]{@{}l@{}}6.83 mm in translation\\5.42 $^{\circ}$ in orientation \end{tabular} \\ \hline
Jun et al. \cite{jun2016calibration}          & 8 $\times$ 1 paired corresponding point & Fingertip & Inside-out depth tracking                                                  & 
\begin{tabular}[c]{@{}l@{}} $\approx$ 1 cm in translation \end{tabular} \\\hline
\end{tabular}
\end{table*}

\begin{figure}[htb]
 \centering 
 \includegraphics[width=\columnwidth]{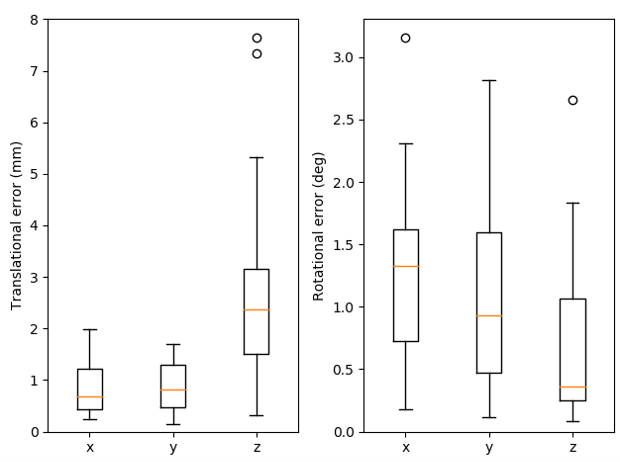}
 \caption{Boxplot of 3D translational and rotational error.}
 \label{fig:boxplot}
\end{figure}

\paragraph{Results} 
As shown in \autoref{tab:result} and \autoref{fig:boxplot}, the virtual-to-real alignment was found to be reliable in both translation and rotation after calibration. The planar positional error equals to 8.81$\pm$2.69 arcmin or 1.37$\pm$0.42 mm. The overall 3D rotational error is 1.76$\pm$0.99$^{\circ}$. As expected, the highest misalignment exists in depth, because the depth perception that relies on relative size change is not as salient as planar contour features \cite{azimi2017alignment}. Due to the same reason, the rotation is most accurate around the z-axis. \autoref{tab:compare} shows a comparison of 3D AR overlay error between our calibration method and some user-centric alignment-based methods. As physical misalignment (in mm) may be biased by the target distance, for a fair comparison, the selected works were all calibrated and evaluated at arm-reach distance. Our method yields a comparable overlay accuracy with the MPAAM-based work by Azimi et al. \cite{azimi2017alignment} and considerably better results than the other two.

\subsection{Assessment with user test}
\begin{figure}[htb]
 \centering 
 \includegraphics[width=1\columnwidth]{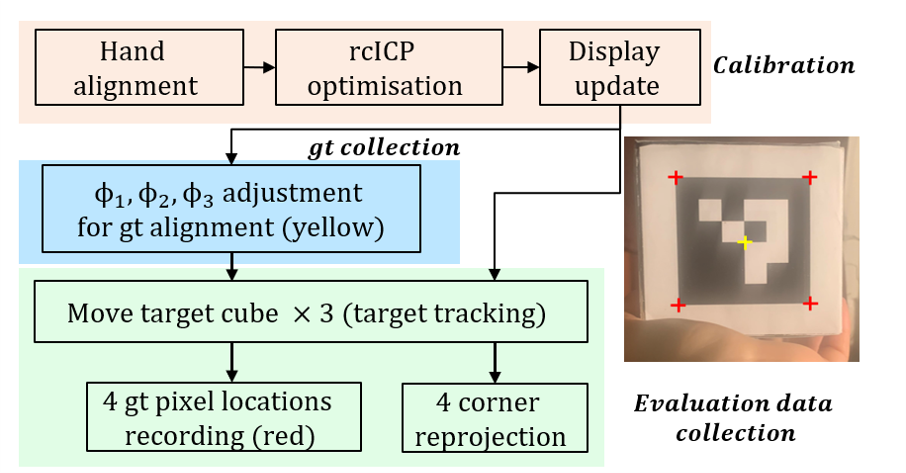}
 \caption{The workflow of user test for each user session.}
 \label{fig:user}
\end{figure}

\paragraph{Test design}
User involvement is essential to define the gt pixel that perfectly aligns with its real counterpart in the retina \cite{mcgarrity2001new,Tang:2003}. Three expert users (two males and one female research students, including one of the authors) familiar with the calibration procedure performed 10 calibration trials each. Note that unlike the camera-based evaluation that was performed on one single display side, in the user test the calibration and the evaluation were based on the stereo observation. The two displays were simultaneously calibrated as mentioned before. As shown in \autoref{fig:user}, after completing the proposed hand-based alignment, each user was asked to manually correct the position of rendered cube by adjusting $\tilde{\phi_1}, \tilde{\phi_2}$ and $\tilde{\phi_3}$ so that the virtual cube aligned with the observed real cube by their centre (yellow cross). Then, the target cube was randomly placed at 0.4m, 0.5m and 0.6m. The pixel locations of four evaluation corners (red crosses) were recorded as $\mathbf{m}_{\scriptscriptstyle{gt}}$ for testing. 

The reprojected pixels $\mathbf{m}_{\scriptscriptstyle{reproj}}$ was computed with the updated projection calibrated by the propsoed method:
\begin{equation}
    \mathbf{m}_{\scriptscriptstyle{reproj}}(t) =  \prescript{\mathrm{on-E}}{}{\mathbf{K}} \mathbf{H}_{\scriptscriptstyle 0}\mathbf{UQ}
\begin{bmatrix}
\mathbf{I}_{\scriptscriptstyle {3\times3}} &\mathbf{t}_{\scriptscriptstyle {C_{\scriptscriptstyle 0}V}}\\
0&1
\end{bmatrix} \prescript{\scriptscriptstyle {V}}{\scriptscriptstyle W}{\mathbf{T}}(t)
~ \mathbf{v}^{\scriptscriptstyle {(W)}}(t)
\end{equation}
Similar to other literature works, the reprojection error was defined as the mean distance between $\mathbf{m}_{\scriptscriptstyle{gt}}$ and $\mathbf{m}_{\scriptscriptstyle{reproj}}$, using all collected $n=$360 (i.e., 3 users$\times$10 trials$\times$3 distances$\times$4 corners) pairs of data: 
\begin{equation}
    err_{\scriptscriptstyle{reproj}} = \frac{1}{n}\sum_{i=1}^{n}\norm{\mathbf{m}_{\scriptscriptstyle{gt,~i}} - \mathbf{m}_{\scriptscriptstyle{reproj,~i}}}
\end{equation}

\begin{figure}[htb]
 \centering 
 \includegraphics[width=1\columnwidth]{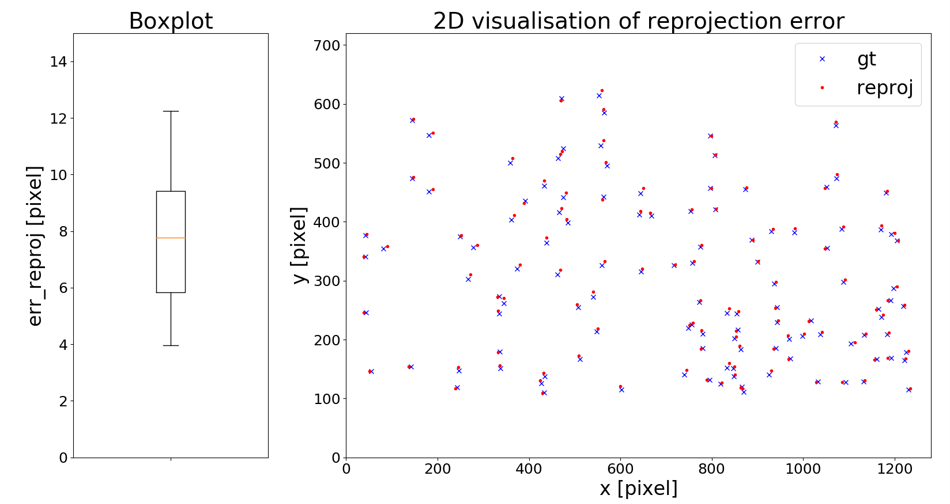}
 \caption{Left: a boxplot of the 2D projection error with the y
axis showing $err_{\scriptscriptstyle{reproj}}$. Right: visualisation of the reprojected and GT points.}
 \label{fig:scatter}
\end{figure}

\paragraph{Results}
As shown in \autoref{fig:scatter}, the calibrated display achieves an error of 7.71$\pm$2.29 pixels which equals to 10.79$\pm$3.21 arcmin according to the HoloLens display average angular resolution. \autoref{tab:reproj} shows a comparison of the 2D reprojection error with some other OST calibration methods regarding median, interquartile range (IQR) and extreme values. Note that the pixel misalignment has been converted into arcmin to avoid the bias by different visors properties (e.g., resolutions and focal distances). Our method achieves a better accuracy than the eye tracking-based calibration by \cite{ItohIndica:2014} and the fingertip-based calibration by \cite{MoserLeapMotion:2016}, but a worse accuracy than the stylus-aligned calibration. 

\begin{table}[bth]
\centering
\caption{Comparison of 2D reprojection error with some OST calibration methods.}
\label{tab:reproj}
\begin{tabular}{p{2.1cm}p{5.5cm}}
\hline
Studies               & Accuracy (arcmin)                     \\ \hline
Ours                  &  Median: 10.87, IQR: 8.18-13.18, Min: 5.55, Max: 17.16
\\ \hline
% SPAAM \cite{ItohIndica:2014}   & Median: 11.76, IQR: 10.25-13.26, Min: 7.11, Max: 16.67 \\ \hline
INDICA \cite{ItohIndica:2014}   & Median: 14.49, IQR: 12.10-16.08, Min: 10.42 Max: 17.32 \\ \hline
Moser et al. \cite{MoserLeapMotion:2016}   &
\begin{tabular}[c]{@{}l@{}}(Stylus) Median: 9.57, IQR: 6.13-14.80,\\ Min: 0.66, Max: 13.09 \\\hline
(Finger) Median: 16.98, IQR: 10.83-25.51,\\ Min: 0.93, Max: 46.87
 \end{tabular}
 \\ \hline
\end{tabular}
\end{table}

\section{Discussion}
In state-of-the-art works, each alignment can provide only one or a few correspondences. Therefore, repetitive or multi-target alignments are essential. By contrast, in our work, each alignment collects numerous unpaired corresponding points. A single alignment can potentially ensure the reliable OST calibration update, thereby reducing the time and workload required for users.

For general MPAAM solutions \cite{azimi2017alignment, GuoOSTCalibrationOnline:2019}, the 3D alignment between the target calibration tool and perceived virtual homologous is unconstrained (i.e., user must align in full 6 DOFs) and totally dependent on the user's judgment for overlay quality. However, the rotational alignment is prone to errors due to the reduced depth perception typical of non light-field OST displays. Since the OST update problem is 3 DOFs (i.e., viewpoint shift) by nature, the additional DOFs in rotation will degrade the calibration accuracy and are thus undesired. In our setup, users are instructed to maintain the hand orientation while translating in 3 DOFs under the computer guidance. Furthermore, the rcICP optimisation can tolerate up to 9$^{\circ}$ rotational misalignment, making the overall calibration more robust. Given a reliable offline calibration, since the initial cursor is generated at the user-specified locations, only a centimeter-level translation is normally required for proper alignment. According to our test, the hand shape can be easily retained during hand translations.

To the best of our knowledge, we are the first that uses the raw RGBD camera output (in hand's surface point cloud) for OST calibration. In two similar hand-exploited works \cite{jun2016calibration, MoserLeapMotion:2016}, the fingertip needed to be additionally computed from the raw samplings. The benefit of depth sensing (i.e., the precise measurement of a continuous target surface), however, was not fully exploited by those sparse fingertip-wise alignment procedures. Furthermore, due to the vague definition of the fingertip in hand anatomy, the tracked fingertip positions were inevitably noisy. The sparse correspondences collected with such noisy landmarks may explain the high error of these two methods in \autoref{tab:compare} and \autoref{tab:reproj}.

% Unlike other existing methods, our method ``solves in 3D, updates in 2D": by considering two 3D-2D rendering pipelines at different time steps, the unknown parameters can be optimised solely in 3D space by point cloud registration. Once $\mathbf{UQ}$ is computed, the projection can be corrected by our update model that considers both extrinsic update $\mathbf{Q}$ and intrinsic update $\mathbf{U}$. Such a procedure allows robust optimisation in black-box fashion and precise updates considering details of an eye-display model. 

% As mentioned previously, during the parameter optimisation, the $\mathbf{Q}$ matrix is dominant due to its higher contribution to the 3D-3D relative transformation $\mathbf{M}$ between the two tracked point clouds. However, considering the 3D-2D projection update, the small numbers in $\mathbf{U}$ (i.e., $\phi_1\tilde{\phi_4}, \phi_2\tilde{\phi_4}, \phi_3\tilde{\phi_4}$) will be amplified by the multiplication by $\prescript{\mathrm{off-E}}{}{\mathbf{K}_{\scriptscriptstyle 0}}$, since the first two numbers in the diagonal of $\prescript{\mathrm{off-E}}{}{\mathbf{K}_{\scriptscriptstyle 0}}$ (i.e., focal length in pixels) are usually large (e.g., around 1560 in our case).
% As suggested by Genc et al. \cite{GencEasySPAAM:2002},  either in-image translations (associates with our $\mathbf{U}$ matrix) or in-space translations (associated with our $\mathbf{Q}$ matrix) can be dominant, depending on the practical situation under consideration \cite{GencEasySPAAM:2002}.

\section{Conclusion}
In this work, we presented an user-centric alignment-based OST calibration method that is robust to rotational misalignment. 
By exploiting a commercial RGBD camera, a large number of unpaired corresponding points can be collected by each object-wise alignment performed with the user's bare-hand. We proposed a novel rcICP method developed in accordance with the updated eye-display pinhole camera model, which aims to optimize the three unknown parameters from the collected unpaired points. According to the simulation tests, such rcICP optimisation is robust against up to 9$^{\circ}$ rotational disturbance that is possibly caused by the imperfect hand alignment performed by unskilled users. With our implementation of the proposed calibration with the HoloLens 1 and at arm-reach distances, 
the camera-based evaluation shows an average 8.81 arcmin (1.37 mm) overlay misalignment, whereas the users test shows an average reprojection error of 10.79 arcmin (7.71 pixels). Results indicate that our method is more accurate than other finger-based methods and comparable with some visually tracked tool-based methods, even when a single bare-hand alignment is performed.

% %% if specified like this the section will be committed in review mode
% \acknowledgments{
% The authors wish to thank A, B, C. This work was supported in part by
% a grant from XYZ.}

\bibliographystyle{abbrv-doi}

\bibliography{template}
\end{document}